\documentclass[12pt]{article}
\begin{document}
\newcommand{\beq}{\begin{equation}}
\newcommand{\eeq}{\end{equation}}
\newcommand{\beqa}{\begin{eqnarray}}
\newcommand{\eeqa}{\end{eqnarray}}
\newcommand{\beqar}{\begin{eqnarray*}}
\newcommand{\eeqar}{\end{eqnarray*}}
\newcommand{\al}{\alpha}
\newcommand{\be}{\beta}
\newcommand{\del}{\delta}
\newcommand{\D}{\Delta}
\newcommand{\eps}{\epsilon}
\newcommand{\ga}{\gamma}
\newcommand{\Ga}{\Gamma}
\newcommand{\ka}{\kappa}
\newcommand{\nn}{\nonumber}
\newcommand{\inn}{\!\cdot\!}
\newcommand{\h}{\eta}
\newcommand{\ii}{\iota}
\newcommand{\kk}{\varphi}
\newcommand\F{{}_3F_2}
\newcommand{\la}{\lambda}
\newcommand{\La}{\Lambda}
\newcommand{\na}{\prt}
\newcommand{\Om}{\Omega}
\newcommand{\om}{\omega}
\newcommand{\p}{\phi}
\newcommand{\sig}{\sigma}
\renewcommand{\t}{\theta}
\newcommand{\z}{\zeta}
\newcommand{\ssc}{\scriptscriptstyle}
\newcommand{\eg}{{\it e.g.,}\ }
\newcommand{\ie}{{\it i.e.,}\ }
\newcommand{\labell}[1]{\label{#1}} 
\newcommand{\reef}[1]{(\ref{#1})}
\newcommand\prt{\partial}
\newcommand\veps{\varepsilon}
\newcommand{\pol}{\varepsilon}
\newcommand\vp{\varphi}
\newcommand\ls{\ell_s}
\newcommand\cF{{\cal F}}
\newcommand\cA{{\cal A}}
\newcommand\cS{{\cal S}}
\newcommand\cT{{\cal T}}
\newcommand\cV{{\cal V}}
\newcommand\cL{{\cal L}}
\newcommand\cM{{\cal M}}
\newcommand\cN{{\cal N}}
\newcommand\cG{{\cal G}}
\newcommand\cH{{\cal H}}
\newcommand\cI{{\cal I}}
\newcommand\cJ{{\cal J}}
\newcommand\cl{{\iota}}
\newcommand\cP{{\cal P}}
\newcommand\cQ{{\cal Q}}
\newcommand\cg{{\it g}}
\newcommand\cR{{\cal R}}
\newcommand\cB{{\cal B}}
\newcommand\cO{{\cal O}}
\newcommand\tcO{{\tilde {{\cal O}}}}
\newcommand\bg{\bar{g}}
\newcommand\bb{\bar{b}}
\newcommand\bH{\bar{H}}
\newcommand\bc{\bar{c}}
\newcommand\bK{\bar{K}}
\newcommand\bF{\bar{F}}
\newcommand\bZ{\bar{Z}}
\newcommand\bxi{\bar{\xi}}
\newcommand\bphi{\bar{\phi}}
\newcommand\bpsi{\bar{\psi}}
\newcommand\bprt{\bar{\prt}}
\newcommand\bet{\bar{\eta}}
\newcommand\btau{\bar{\tau}}
\newcommand\hF{\hat{F}}
\newcommand\hA{\hat{A}}
\newcommand\hT{\hat{T}}
\newcommand\htau{\hat{\tau}}
\newcommand\hD{\hat{D}}
\newcommand\hf{\hat{f}}
\newcommand\hg{\hat{g}}
\newcommand\hp{\hat{\phi}}
\newcommand\hi{\hat{i}}
\newcommand\ha{\hat{a}}
\newcommand\hb{\hat{b}}
\newcommand\hQ{\hat{Q}}
\newcommand\hP{\hat{\Phi}}
\newcommand\hS{\hat{S}}
\newcommand\hX{\hat{X}}
\newcommand\tL{\tilde{\cal L}}
\newcommand\hL{\hat{\cal L}}
\newcommand\tG{{\widetilde G}}
\newcommand\tg{{\widetilde g}}
\newcommand\tphi{{\widetilde \phi}}
\newcommand\tPhi{{\widetilde \Phi}}
\newcommand\te{{\tilde e}}
\newcommand\tk{{\tilde k}}
\newcommand\tf{{\tilde f}}
\newcommand\ta{{\tilde a}}
\newcommand\tb{{\tilde b}}
\newcommand\tR{{\tilde R}}
\newcommand\teta{{\tilde \eta}}
\newcommand\tF{{\widetilde F}}
\newcommand\tK{{\widetilde K}}
\newcommand\tE{{\widetilde E}}
\newcommand\tpsi{{\tilde \psi}}
\newcommand\tX{{\widetilde X}}
\newcommand\tD{{\widetilde D}}
\newcommand\tO{{\widetilde O}}
\newcommand\tS{{\tilde S}}
\newcommand\tB{{\widetilde B}}
\newcommand\tA{{\widetilde A}}
\newcommand\tT{{\widetilde T}}
\newcommand\tC{{\widetilde C}}
\newcommand\tV{{\widetilde V}}
\newcommand\thF{{\widetilde {\hat {F}}}}
\newcommand\Tr{{\rm Tr}}
\newcommand\tr{{\rm tr}}
\newcommand\STr{{\rm STr}}
\newcommand\hR{\hat{R}}
\newcommand\M[2]{M^{#1}{}_{#2}}

\newcommand\bS{\textbf{ S}}
\newcommand\bI{\textbf{ I}}
\newcommand\bJ{\textbf{ J}}

\begin{titlepage}
\begin{center}

\vskip 2 cm
{\LARGE \bf   T-duality constraint on R-R couplings
 }\\
\vskip 1.25 cm
   Mohammad R. Garousi\footnote{garousi@um.ac.ir}

\vskip 1 cm
{{\it Department of Physics, Faculty of Science, Ferdowsi University of Mashhad\\}{\it P.O. Box 1436, Mashhad, Iran}\\}
\vskip .1 cm
 \end{center}

\begin{abstract}
 It has been speculated   that the metric, $B$-field and dilaton couplings in the  low energy effective action of   string theory at any order of $\alpha'$ may be found  by imposing the gauge symmetries and  the T-duality  on the effective action.  This proposal can be extended to  include  the  Ramond-Ramond (R-R)  couplings as well. In this paper, we first find the dimensional reduction of the R-R fields  and then  perform explicitly the T-duality constraint on the R-R couplings   at the supergravity  level. Up to an overall factor, it reproduces  the democratic form of the R-R couplings.
 
 \vskip 2 cm
  
  
\end{abstract}
\end{titlepage}

One of the most exciting discoveries in perturbative  string theory is   T-duality \cite{Giveon:1994fu,Alvarez:1994dn,Becker:2007zj} which appears  when one compactifies theory  on a torus. It has been speculated  that the invariance of the effective action of bosonic string theory under the standard gauge transformations  and under non-standard  T-duality transformations which receive $\alpha'$-corrections,   may be used as  constraints to construct the  low energy effective action  of the string theory  \cite{Garousi:2017fbe}. In this approach, using the field redefinitions freedom \cite{Metsaev:1987zx},  
one first constructs the minimum number of  gauge invariant couplings. Then one reduces them  on   a circle. The reduced actions must be invariant under   the standard Buscher rules  \cite{Buscher:1987sk,Buscher:1987qj} plus their  $\alpha'$-corrections  \cite{Tseytlin:1991wr,Bergshoeff:1995cg,Kaloper:1997ux,Garousi:2019wgz}.       Using this approach, the effective action of the bosonic string theory up to order $\alpha'^3$  have been found in \cite{Garousi:2019wgz,Garousi:2019mca}.  The NS-NS part of the effective action of the superstring  theory on a manifold with boundary at the leading order of $\alpha'$ has been also found in \cite{Garousi:2019xlf} by imposing gauge symmetries, the T-duality and by imposing the S-duality, another duality which exists in the superstring theory \cite{Becker:2007zj}. In particular, the well known Gibbons-Hawking-York term \cite{York:1972sj,Gibbons:1976ue} has been dictated   by the string dualities    \cite{Garousi:2019xlf}.

Another  T-duality based approach for constructing  the $D$-dimensional effective action  is the   Double Field Theory      \cite{Siegel:1993xq,Siegel:1993th,Siegel:1993bj,Hull:2009mi,Aldazabal:2013sca} in which the effective action in $2D$-space  is constrained  to be invariant under T-duality and under  gauge transformations. The T-duality in this case, however,   is  the standard  $O(D,D)$ transformation without $\alpha'$-corrections whereas the gauge transformation is non-standard  which  receives   $\alpha'$-corrections \cite{Hohm:2010pp,Aldazabal:2013sca,Hohm:2014xsa,Marques:2015vua}.  This approach has been extended in \cite{Hohm:2011dv}  to type II superstring theories. These T-duality approaches for constructing the effective actions are based on the observation made by Sen in the context of closed string field theory \cite{Sen:1991zi} that the effective action of string theory should be invariant under T-duality to all orders in $\alpha'$.

In this paper we would like to extend the first approach  to the couplings in type II  superstring theories.  The type II string theories have  NS-NS fields which are the same as the fields in the bosonic string theory, as well as some  R-R fields  which are  also bononic fields. The $D_p$-branes in type II string theories carry the R-R charges \cite{Polchinski:1995mt}. These theories have also NS-R and R-NS  femionic fields in which we are not interested. The odd-form R-R potentials appear in type IIA and even-forms appear in type IIB.   It is known that the compactification of type IIA theory on a circle transforms to the compactification of  type IIB theory  on  another circle under the T-duality transformations. To study the effective action of the bosonic fields in  these theories, it is convenient  to collect the two theories to one theory which is called type II theory. It has both odd- and even-form R-R potentials.  When comactifying this theory on a   circle,  the effective action then is expected to be invariant under the T-duality transformations, as in the bosonic string theory.  

When compactifying string theory on a circle with unit radius and with the killing coordinate $y$, the T-duality transformations for the NS-NS fields   are the Buscher rules \cite{Buscher:1987sk,Buscher:1987qj}, \ie
\beqa
e^{2\phi'}=\frac{e^{2\phi}}{G_{yy}}&;& 
G'_{yy}=\frac{1}{G_{yy}}\nonumber\\
G'_{\mu y}=\frac{B_{\mu y}}{G_{yy}}&;&
G'_{\mu\nu}=G_{\mu\nu}-\frac{G_{\mu y}G_{\nu y}-B_{\mu y}B_{\nu y}}{G_{yy}}\nonumber\\
B'_{\mu y}=\frac{G_{\mu y}}{G_{yy}}&;&
B'_{\mu\nu}=B_{\mu\nu}-\frac{B_{\mu y}G_{\nu y}-G_{\mu y}B_{\nu y}}{G_{yy}}\labell{nonlinear}
 \eeqa
where $\mu,\nu$ denote any   direction other than $y$. In above transformations the metric is  in the string frame.  If one assumes fields are transformed covariantly under the coordinate transformations, then the above transformations receive $\alpha'$-corrections \cite{Tseytlin:1991wr,Bergshoeff:1995cg,Kaloper:1997ux,Garousi:2019wgz}.
The T-duality transformations of the R-R fields at the leading order of $\alpha'$ have been found in \cite{Meessen:1998qm}, \ie
\beqa
C'^{(n)}_{\mu\cdots \nu \alpha y}&=& C^{(n-1)}_{\mu\cdots \nu \alpha}-\frac{C^{(n-1)}_{[\mu\cdots\nu|y}G_{|\alpha]y}}{G_{yy}}\labell{C'}\\
C'^{(n)}_{\mu\cdots\nu\alpha\beta}&=&C^{(n+1)}_{\mu\cdots\nu\alpha\beta y}+C^{(n-1)}_{[\mu\cdots\nu\alpha}B_{\beta]y}+\frac{C^{(n-1)}_{[\mu\cdots\nu|y}B_{|\alpha|y}G_{|\beta]y}}{G_{yy}}\nn
\eeqa
They may  also have $\alpha'$ corrections in which we are not interested in this paper. Our notation for making  antisymmetry  is such that \eg $C^{(2)}_{[\mu_1\mu_2}B_{\mu_3]y}=C^{(2)}_{\mu_1\mu_2}B_{\mu_3y}+C^{(2)}_{\mu_2\mu_3}B_{\mu_1y}-C^{(2)}_{\mu_1\mu_3}B_{\mu_2y}$. The T-duality transformations \reef{nonlinear} and \reef{C'} are such that they are consistent with the fact that $D_p$-brane in type II theory transform to $D_{p-1}$-brane or $D_{p+1}$-brane depending on whether the brane is along or orthogonal to the circle on which the T-duality is imposed. In fact the R-R fields couple to the $D_p$-brane  as  
\beqa
\int_{M^{p+1}}e^BC\labell{BC}
\eeqa
where $C=\sum_{n=0}^{8}C^{(n)}$. It is invariant under the R-R gauge transformation $\delta C=d\Lambda+H\Lambda$ where $\Lambda=\sum_{n=0}^7\Lambda^{(n)}$. The T-duality transformations \reef{nonlinear} and \reef{C'} produce the following  transformations:
\beqa
(e^BC)'_{\cdots y} \,=\,(e^BC)_{\cdots}&;&(e^BC)'_{\cdots } \,=\,(e^BC)_{\cdots y}
\eeqa
where dots represent some world-volume indices. In other words, the coupling \reef{BC} is covariant under the T-duality transformations.

 The effective action of type II string theory on the closed manifolds at the leading order of $\alpha'$  is the well-known type II supergravity (see \eg \cite{Becker:2007zj}). The first higher derivative corrections to this action is at order $\alpha'^3$. The Riemann curvature couplings at this order are known in the literature \cite{Gross:1986iv,Grisaru:1986vi,Freeman:1986zh}. There are many other couplings involving B-field, dilaton and R-R fields at this order.   Some of them have been found in \cite{Policastro:2008hg, Liu:2013dna,Minasian:2015bxa}.  There are also boundary terms at this order when manifolds have boundary which are not known in the literature. We expect all these  couplings might be found by imposing the appropriate gauge transformations and string duality constraints on the effective action. In fact, the known Riemann curvature couplings are reproduced by this method in \cite{Razaghian:2018svg}. In this paper we present the technical  details for imposing the T-duality constraint to reproduce the R-R couplings at the supergravity level and leave the calculations at order $\alpha'^3$ for the future works. The NS-NS couplings at the leading order of $\alpha'$ are the same as the corresponding couplings in the bosonic theory.

The R-R couplings, as in \reef{BC}, should be invariant under the R-R gauge transformations. The R-R couplings in the effective action  should be in terms of the R-R field strength, \ie
\beqa
F^{(n)}&=&dC^{(n-1)}+H\wedge C^{(n-3)}\labell{Fn0}
\eeqa
which is invariant under the R-R gauge transformations\footnote{The definition of the R-R field strength for $n=5$ is slightly different from the common definition in the supergravity literature  (see \eg \cite{Becker:2007zj}). However, the field redefinition of the R-R potentials $C^{(4)}\rightarrow C^{(4)}+\frac{1}{2}B\wedge C^{(2)}$ and $C^{(2)}\rightarrow -C^{(2)}$ transforms the above $F^{(5)}$ to the standard form of $F^{(5)}=dC^{(4)}-\frac{1}{2}H\wedge C^{(2)}+\frac{1}{2}B\wedge dC^{(2)}$.}. Since $F^{(n)}$ is not the exterior derivative of an $(n-1)$-form for $n>2$, it satisfies the following anomalous Bianchi identity:
\beqa
dF^{(n)}&=&-H\wedge F^{(n-2)}\labell{Bian}
\eeqa
At the two-derivative level, the gauge invariance constraint then requires the following couplings in the string frame:
\beqa
\bS_{0}^{RR}&=& -\frac{2}{\kappa^2}\int d^{10}x \sqrt{-G}\,  \sum_{n=1}^{9}a_n|F^{(n)}|_G^2\labell{S0b}\\
&=& -\frac{2}{\kappa^2}\int d^{10}x \sqrt{-G}\,  \sum_{n=1}^{9}\frac{a_n}{n!}G^{a_1b_1}\cdots G^{a_nb_n}F^{(n)}_{a_1\cdots a_n}F^{(n)}_{b_1\cdots b_n}\nn
\eeqa
where $a_1,a_2,\cdots,a_{9}$ are 9 parameters that the R-R gauge symmetry can not fix them. Since the R-R field strengths \reef{Fn0} are nonlinear for $n>2$, these constants can not be absorbed by the normalization of the R-R potentials. We are going to show that they all can  be written in terms of $a_1$ by imposing on them the T-duality constraint as well. We did not include $F^{(0)}$ and $F^{(10)}$ terms in above couplings because they  do not include dynamical fields.  In writing the above couplings we assume the R-R fields are all independent. The on-shell physics, however, requires not all components of $C^{(4)}$ to be independent. Moreover, the fields $ C^{(5)},C^{(6)}, C^{(7)},C^{(8)}$ are not  independent. That means in the equations of motion one has to impose some extra constraints on the R-R field strengths to have physical fields in  the  equations of motion.

To impose the T-duality constraint on the effective action \reef{S0b}, we have to consider the Kaluza-Klein reduction of the 10-dimensional metric, \ie
\beqa
ds^2&=&\bg_{\mu\nu}dx^\mu dx^\nu+e^{\varphi}(dy+g_\mu dx^\mu)^2
\eeqa
which is invariant  under  $y$-coordinate transformation $y\rightarrow y-\lambda(x)$ and the gauge transformation $g_\mu\rightarrow g_\mu+\prt_\mu\lambda$. Apart from this $U(1)$ isometry gauge symmetry,  the reduced theory should have another $U(1)$ gauge symmetry corresponding to the Kalb-Ramond form $B=\frac{1}{2}B_{\mu\nu}dx^\mu\wedge dx^\nu+B_{\mu y}dx^\mu\wedge dy$. The reduction of any theory containing  the B-field is  manifestly the $U(1)\times U(1)$ gauge invariant  if one uses the following reduction for B-field and dilaton as well  \cite{Maharana:1992my}:
  \beqa
G_{ab}=\left(\matrix{\bg_{\mu\nu}+e^{\varphi}g_{\mu }g_{\nu }& e^{\varphi}g_{\mu }&\cr e^{\varphi}g_{\nu }&e^{\varphi}&}\right),\, B_{ab}= \left(\matrix{\bb_{\mu\nu}+\frac{1}{2}b_{[\mu }g_{\nu] }&b_{\mu }\cr - b_{\nu }&0&}\right),\,  \Phi=\bar{\phi}+\varphi/4\labell{reduc}\eeqa
where $\bg_{\mu\nu}, \bb_{\mu\nu}, \bar{\phi} $ are the metric,  a two-form and the dilaton, respectively,   in the base space, and $g_{\mu},\, b_{\mu}$ are two vectors  in this space. Inverse of the above $10$-dimensional metric is 
\beqa
G^{ab}=\left(\matrix{\bg^{\mu\nu} &  -g^{\mu }&\cr -g^{\nu }&e^{-\varphi}+g_{\alpha}g^{\alpha}&}\right)\labell{inver}
\eeqa
where $\bg^{\mu\nu}$ is the inverse of the base space  metric which raises the index of the   vectors. The Buscher rules \reef{nonlinear}  in the parametrizations \reef{reduc}  become  the following linear transformations:
\beqa
\varphi'= -\varphi
\,\,\,,\,\,g'_{\mu }= b_{\mu },\,\, b'_{\mu }= g_{\mu } ,\,\,\bg_{\alpha\beta}'=\bg_{\alpha\beta} ,\,\,\bb_{\alpha\beta}'=\bb_{\alpha\beta} ,\,\,  \bar{\phi}'= \bar{\phi}\labell{T21}
\eeqa
They form a $Z_2$-group, \ie $ (x')'= x$ where $x$ is any field in the base space. These transformations receive higher derivative corrections in which we are not interested in this paper.

The reduction of field strength of B-field  in the parametrizations \reef{reduc} becomes
\beqa
H_{\mu\nu\alpha}&=& \bH_{\mu\nu\alpha}+g_{[\mu}W_{\nu\alpha]}\nn\\
H_{\mu\nu y}&=&W_{\mu\nu}\labell{rH}
\eeqa
where   $W$ is field strength of the $U(1)$ gauge field $b_{\mu}$, \ie $W=db$, and the three-form $\bH$ which is torsion in the base space, is defined as
\beqa
 \bH_{\mu\nu\alpha}&\equiv &\hat{H}_{\mu\nu\alpha}-\frac{1}{2}g_{[\mu} W_{\nu\alpha]}-\frac{1}{2}b_{[\mu} V_{\nu\alpha]}
 \eeqa
 where $\hat{H}$ is field strength of the two-form $\bb_{\mu\nu}$ and  $V$ is field strength of the $U(1)$ gauge field $g_{\mu}$, \ie $V=dg$.  The three-form $\bH$ is invariant under the T-duality and under various  gauge transformations, \eg under the $U(1)$ isometry gauge transformation,  the $B_{\mu\nu}$ components of B-field transform as $B_{\mu\nu}\rightarrow B_{\mu\nu}+b_{\mu}\prt_{\nu}\lambda-b_{\nu}\prt_{\mu}\lambda$. Hence, $H_{\mu\nu\alpha}\rightarrow H_{\mu\nu\alpha}+\prt_{[\mu}\lambda W_{\nu\alpha]}$. The anomalous term is cancelled with the transformation of the last term in the first line of \reef{rH}. Hence, $\bH_{\mu\nu\alpha}$ remains invariant.  It is also obvious from \reef{rH} that $\bH$ is invariant under the B-field gauge transformation $B_{ab}\rightarrow B_{ab}+\prt_a\omega_b-\prt_b\omega_a$. Since $\bH$ is not exterior derivative of a two-form,  it satisfies the following anomalous Bianchi identity:
 \beqa
 \prt_{[\mu} \bH_{\nu\alpha\beta]}&=&-V_{[\mu\nu}W_{\alpha\beta]}\labell{anB}
 \eeqa
  Using the reduction \reef{reduc}, one finds that the reduction of any 10-dimensional gauge invariant coupling  can be written in terms of gauge invariant tensors $\bar{R},\bH,V,W, \bphi,\varphi$ and their derivatives. Hence, the reduction is consistent with the  $U(1)\times U(1)$ gauge symmetry \cite{Maharana:1992my}. In this paper, among other things, we are going  to find such reduction for the R-R fields.

To find such reduction for the R-R fields, we first note that  in the parametrization \reef{reduc} the non-linear T-duality transformations of the R-R fields \reef{C'}  become
\beqa
C'^{(n)}_{\mu\cdots \nu \alpha y}&=& C^{(n-1)}_{\mu\cdots \nu \alpha}-C^{(n-1)}_{[\mu\cdots\nu|y}g_{\alpha]} \labell{C'1}\\
C'^{(n)}_{\mu\cdots\nu\alpha\beta}&=&C^{(n+1)}_{\mu\cdots\nu\alpha\beta y}+C^{(n-1)}_{[\mu\cdots\nu\alpha}b_{\beta]}+ C^{(n-1)}_{[\mu\cdots\nu|y}b_{\alpha}g_{\beta]} \nn
\eeqa
which are still nonlinear. On the right-hand side the R-R fields are 10-dimensional whereas the $b_{\alpha},g_{\alpha}$ are 9-dimensional fields. To proceed further then  one has to reduce the R-R fields as well. The reduction should be consistent with the  $U(1)\times U(1)$ gauge symmetry.

We consider 
    the following reductions for the R-R fields:
\beqa
C^{(n)}_{\mu_1\cdots\mu_n}&=&\bc^{(n)}_{\mu_1\cdots\mu_n}+\bc^{(n-1)}_{[\mu_1\cdots\mu_{n-1}}g_{\mu_n]}\nn\\
C^{(n+1)}_{\mu_1\cdots \mu_ny}&=&\bc^{(n)}_{\mu_1\cdots\mu_n}\labell{CC}
\eeqa
where $\bc^{(n)}$ are R-R potentials in the $9$-dimensional base space. 
The nonlinear T-duality transformations  \reef{C'} in the above  parametrizations become  the following linear transformations:
\beqa
\bc'^{(n)}_{\mu_1\cdots\mu_n}=\bc^{(n)}_{\mu_1\cdots\mu_n}\labell{T2}
\eeqa
It is remarkable that  in the   parametrizations \reef{CC}, the 9-dimensional R-R fields become invariant under the  T-duality transformation. They may, however, receive corrections at the higher order of $\alpha'$ in which we are not interested in this paper.

The reduction of the 10-dimensional R-R field strength  in the   parametrizations \reef{CC} becomes 
\beqa
F^{(n)}_{\mu_{1}...\mu_{n}}&=& {F^V}^{(n)}_{\mu_{1}...\mu_{n}}+g_{[\mu_1}F^{W(n-1)}_{\mu_2\cdots\mu_n]}\nn\\
F^{(n)}_{\mu_{1}...\mu_{n-1}y}&=& {F^W}^{(n-1)}_{\mu_{1}...\mu_{n-1}}\labell{FF}
\eeqa
The   forms $ {F^W}^{(n)}$ and  $ {F^V}^{(n)}$  are defined as 
\beqa
 {F^W}^{(n)}&\equiv & \bF^{(n)}+\bH\wedge \bc^{(n-3)}+(-1)^{(n-2)} \bc^{(n-2)}\wedge W\nn\\
 {F^V}^{(n)}&\equiv & \bF^{(n)}+\bH \wedge \bc^{(n-3)}+(-1)^{(n-2)}  \bc^{(n-2)}\wedge V\labell{FWV}
 \eeqa
 where  $\bF^{(n)}$ is field strength of the $9$-dimensional R-R potential $\bc^{(n-1)}$, \ie $\bF^{(n)}=d\bc^{(n-1)}$. These forms are   invariant  under various  gauge transformations, \eg under the $U(1)$ isometry gauge transformation,  the $C^{(n)}_{\mu_1\cdots\mu_n}$ components of the R-R $n$-form potential $C^{(n)}=\frac{1}{n}C^{(n)}_{\mu_1\cdots\mu_n}dx^{\mu_1}\cdots\wedge dx^{\mu_n}+C^{(n)}_{\mu_1\cdots\mu_{n-1}y}dx^{\mu_1}\cdots\wedge dx^{\mu_{n-1}}\wedge dy$ transform as $C^{(n)}_{\mu_1\cdots\mu_n}\rightarrow C^{(n)}_{\mu_1\cdots\mu_n}+C^{(n-1)}_{[\mu_1\cdots\mu_{n-1}}\prt_{\mu_n]}\lambda$. Then one can show that  $F^{(n)}_{\mu_1\cdots\mu_n}\rightarrow F^{(n)}_{\mu_1\cdots\mu_n}+ \prt_{[\mu_1}\lambda F^{W(n-1)}_{\mu_2\cdots\mu_n]}$ where we have  used the fact that $g\wedge g$ and its gauge transformation are zero.  The anomalous term is cancelled with the transformation of the last term in the first line of \reef{FF}. Therefore,  $F^V$ is invariant under the $U(1)$ isometry gauge transformation. Since  $C^{(n)}_{\mu_1\cdots\mu_{n-1}y}$ components of the R-R potential are invariant under the $U(1)$ gauge transformation, the second relation in \reef{FF} indicates that $F^W$ is also invariant. It is also obvious from \reef{FF} that $F^V,F^W$ are invariant under the R-R  gauge transformations because the left-hand sides are invariant. Since the gauge invariant  $n$-forms $F^{V(n)}, F^{W(n)}$ are not exterior derivative  of any $(n-1)$-form, their corresponding Bianchi identities are anomalous.  The anomalous Bianchi identities  are
 \beqa
 dF^{V(n)}&=&(-1)^{(n-2)}F^{W(n-1)}\wedge V-\bH\wedge F^{V(n-2)}\nn\\
  dF^{W(n)}&=&(-1)^{(n-2)}F^{V(n-1)}\wedge W-\bH\wedge F^{W(n-2)}
 \eeqa
 where we have also used the anomalous Bianchi identity \reef{anB} and the fact that $\bH\wedge \bH=0$. Under the T-duality, the gauge invariant forms \reef{FWV} transform as 
\beqa 
 {F^W}^{(n)}&\leftrightarrow &  {F^V}^{(n)}\labell{TFF}
 \eeqa
One expects the reduction of any 10-dimensional coupling involving the R-R field strength and its higher derivatives should be  written in terms of 9-dimensional gauge invariant fields $\bar{R},\bH,V,W, F^{V(n)}, F^{W(n)}, \bphi,\varphi$  and their covariant derivatives.

Using the reductions \reef{reduc}, \reef{inver} and \reef{CC}, it is straightforward to reduce different terms in \reef{S0b}. The reduction of $\sqrt{-G}$ and the R-R coupling $|F^{(1)}|_G^2$,  $|F^{(2)}|_G^2$ and $|F^{(3)}|_G^2$ are the following: 
\beqa
\sqrt{-G}&=&e^{\varphi/2}\sqrt{-\bg} \labell{R}\\
|F^{(1)}|_G^2&=&e^{-\varphi/2}\Big(e^{\varphi/2}|\bF^{(1)}|_{\bg}^2\Big)\nn\\
|F^{(2)}|_G^2&=&e^{-\varphi/2}\Big(e^{-\varphi/2}|\bF^{(1)}|_{\bg}^2+e^{\varphi/2}|\bF^{(2)}+\bc^{(0)}V)|_{\bg}^2\Big)\nn\\
|F^{(3)}|_G^2&=&e^{-\varphi/2}\Big(e^{-\varphi/2} |\bF^{(2)}+ \bc^{(0)}W|_{\bg}^2+e^{\varphi/2}|\bF^{(3)}+\bH \bc^{(0)}-  \bc^{(1)}\wedge V|_{\bg}^2\Big)\nn
\eeqa
The subscript $\bg$ in $|\cdots|^2_{\bg}$ indicates that the indices are contracted with the base space metric $\bg^{\mu\nu}$. The reduction of $|F^{(n)}|_G^2$ for $n>3$ can be written as
 \beqa
 |F^{(n)}|_G^2&=&e^{-\varphi/2}\Big(e^{-\varphi/2}|F^{W(n-1)}|_{\bg}^2+e^{\varphi/2}|F^{V(n)}|_{\bg}^2\Big)\labell{Fn}
\eeqa
As expected, the non-gauge invariant term in \reef{FF} is cancelled in the reduction of 10-dimensional couplings involving the R-R field strength. However, it has gauge invariant contribution to the reduction of  couplings which involve derivatives of the R-R field strength in which we are not interested in this paper.

 Using the fact that  the non-dynamical field strength $F^{(10)}$ in the $10$-dimensional spacetime has been ignored, one should also  ignore the non-dynamical fields in the $9$-dimensional base space. Hence the reduction of  $|F^{(9)}|_G^2$ becomes
 \beqa
 |F^{(9)}|_G^2&=&e^{-\varphi}|F^{W(8)}|_{\bg}^2\labell{F9}
 \eeqa

Having found the reduction of the R-R fields, we now impose the T-duality constraint on the effective action   \reef{S0b}
    to fix the parameters $a_1,\cdots, a_9$. According to this proposal, the effective action should satisfy the following relation:
 \beqa
 S_{\rm eff}(\psi)-S_{\rm eff}(\psi')&=&{\rm TD} \labell{TS}
 \eeqa
where $S_{\rm eff}$ is the  reduction of the $10$-dimensional action on the circle, $\psi$ represents all  massless fields in the base space   and $\psi'$ represents  their transformations under the T-duality  transformations \reef{T21} and \reef{T2}.    On the right-hand side, TD represents some  total derivative terms  in the base space which become zero if  the base space has no boundary. They should be reproduced by the boundary action if the base space has boundary \cite{Garousi:2019xlf}. 

Using the reduction \reef{Fn}, one finds the reduction of  the effective action \reef{S0b} becomes
\beqa
 S^{RR}_{\rm eff}(\psi)&=&-\frac{4\pi}{\kappa^2}\int d^9x\sqrt{-\bg}\Bigg[\sum_{n=1}^{8}a_ne^{\varphi/2}|{F^V}^{(n)}|^2_{\bg}+\sum_{n=1}^{8}a_{n+1}e^{-\varphi/2}|{F^W}^{(n)}|^2_{\bg}\Bigg]\labell{eff1}
\eeqa
Under the T-duality transformations \reef{T21} and \reef{TFF}, it becomes
\beqa
 S^{RR}_{\rm eff}(\psi')&=&-\frac{4\pi}{\kappa^2}\int d^9x\sqrt{-\bg}\Bigg[\sum_{n=1}^{8}a_ne^{-\varphi/2}|{F^W}^{(n)}|^2_{\bg}+\sum_{n=1}^{8}a_{n+1}e^{\varphi/2}|{F^V}^{(n)}|^2_{\bg}\Bigg]\labell{eff2}
\eeqa
One can easily observe that the effective action \reef{S0b} satisfies the constraint \reef{TS} with zero total derivative terms on the right-hand side provided that there is the following recursion relation between the parameters 
\beqa
a_n&=&a_{n+1}\labell{aa}
\eeqa
Hence the T-duality constraint fixes all 9 parameters in \reef{S0b} in terms of one normalization parameter $a_1$. Since there is no total derivative terms on the right-hand side of the T-duality constraint \reef{TS} in this case, the boundary action has no R-R couplings at the leading order of $\alpha'$, as expected. The T-duality constraint on the NS-NS couplings, however, satisfies the relation \reef{TS} with some total derivative terms on the right-hand side which can be cancelled by  the Gibbons-Hawking-York boundary term as well as another boundary term which is not consistent with the S-duality \cite{Garousi:2019xlf}. The duality constraints on the NS-NS couplings also reproduce the standard bulk couplings \cite{Garousi:2019xlf}.

Therefore, the gauge symmetry and the T-duality  fix the low energy effective action of type II string theory in closed spacetime manifold to be 
\beqa
\bS_0&=&  -\frac{2}{\kappa^2}\int d^{10}x \sqrt{-G} \Big[e^{-2\Phi} \left(  R + 4\nabla_{a}\Phi \nabla^{a}\Phi-\frac{1}{12}H^2\right)+ a_1\sum_{n=1}^{9}|F^{(n)}|_G^2\Big]\labell{S0bf}
\eeqa
Up to  the overall factor $a_1$, the R-R couplings are the democratic  R-R couplings that have been found in \cite{Fukuma:1999jt,Hohm:2011dv}. The parameter $a_1$ can  be absorbed by the normalization of the R-R potentials. When spacetime has boundary, the duality constraint dictates that the  Gibbons-Hawking-York boundary term must be also added to the above action. We expect the $\alpha'^3$ corrections to  the action \reef{S0bf} can also be found by imposing on the effective action the gauge symmetries as well as the string dualities. We leave the details of these calculations  for the future works.

 \vskip .3 cm
{\bf Acknowledgments}:   This work is supported by Ferdowsi University of Mashhad.



\begin{thebibliography}{9}


\bibitem{Giveon:1994fu} 
  A.~Giveon, M.~Porrati and E.~Rabinovici,
  Phys.\ Rept.\  {\bf 244}, 77 (1994)
  doi:10.1016/0370-1573(94)90070-1
  [hep-th/9401139].
\bibitem{Alvarez:1994dn} 
  E.~Alvarez, L.~Alvarez-Gaume and Y.~Lozano,
  Nucl.\ Phys.\ Proc.\ Suppl.\  {\bf 41}, 1 (1995)
  doi:10.1016/0920-5632(95)00429-D
  [hep-th/9410237].
  
\bibitem{Becker:2007zj} 
  K.~Becker, M.~Becker and J.~H.~Schwarz,
  ``String theory and M-theory: A modern introduction,''

	
\bibitem{Garousi:2017fbe} 
  M.~R.~Garousi,
  Phys.\ Rept.\  {\bf 702}, 1 (2017)
  doi:10.1016/j.physrep.2017.07.009
  [arXiv:1702.00191 [hep-th]].
	
\bibitem{Garousi:2019xlf} 
  M.~R.~Garousi,
  ``Surface terms in the effective actions via T-duality constraint,''
  arXiv:1907.09168 [hep-th].
\bibitem{Metsaev:1987zx} 
  R.~R.~Metsaev and A.~A.~Tseytlin,
  Nucl.\ Phys.\ B {\bf 293}, 385 (1987).
  doi:10.1016/0550-3213(87)90077-0
  
  
\bibitem{Buscher:1987sk} 
  T.~H.~Buscher,
  Phys.\ Lett.\ B {\bf 194}, 59 (1987).
  doi:10.1016/0370-2693(87)90769-6
\bibitem{Buscher:1987qj} 
  T.~H.~Buscher,
  Phys.\ Lett.\ B {\bf 201}, 466 (1988).
  doi:10.1016/0370-2693(88)90602-8
  
  
\bibitem{Tseytlin:1991wr} 
  A.~A.~Tseytlin,
  Mod.\ Phys.\ Lett.\ A {\bf 6}, 1721 (1991).
  doi:10.1142/S021773239100186X
	
\bibitem{Bergshoeff:1995cg} 
  E.~Bergshoeff, B.~Janssen and T.~Ortin,
  Class.\ Quant.\ Grav.\  {\bf 13}, 321 (1996)
  doi:10.1088/0264-9381/13/3/002
  [hep-th/9506156].
	
\bibitem{Kaloper:1997ux} 
  N.~Kaloper and K.~A.~Meissner,
  Phys.\ Rev.\ D {\bf 56}, 7940 (1997)
  doi:10.1103/PhysRevD.56.7940
  [hep-th/9705193].
  
\bibitem{Garousi:2019wgz} 
  M.~R.~Garousi,
  Phys.\ Rev.\ D {\bf 99}, no. 12, 126005 (2019)
  doi:10.1103/PhysRevD.99.126005
  [arXiv:1904.11282 [hep-th]].
\bibitem{Garousi:2019mca}
M.~R.~Garousi,
Eur. Phys. J. C \textbf{79}, no.10, 827 (2019)
doi:10.1140/epjc/s10052-019-7357-4
[arXiv:1907.06500 [hep-th]].


 
\bibitem{York:1972sj} 
  J.~W.~York, Jr.,
  Phys.\ Rev.\ Lett.\  {\bf 28}, 1082 (1972).
  doi:10.1103/PhysRevLett.28.1082
	 
\bibitem{Gibbons:1976ue} 
  G.~W.~Gibbons and S.~W.~Hawking,
  Phys.\ Rev.\ D {\bf 15}, 2752 (1977).
  doi:10.1103/PhysRevD.15.2752
	
\bibitem{Siegel:1993xq} 
  W.~Siegel,
  Phys.\ Rev.\ D {\bf 47}, 5453 (1993)
  doi:10.1103/PhysRevD.47.5453
  [hep-th/9302036].

\bibitem{Siegel:1993th} 
  W.~Siegel,
  Phys.\ Rev.\ D {\bf 48}, 2826 (1993)
  doi:10.1103/PhysRevD.48.2826
  [hep-th/9305073].
	
\bibitem{Siegel:1993bj} 
  W.~Siegel,
  ``Manifest duality in low-energy superstrings,''
  hep-th/9308133.
	
	
	
\bibitem{Hull:2009mi} 
  C.~Hull and B.~Zwiebach,
  JHEP {\bf 0909}, 099 (2009)
  doi:10.1088/1126-6708/2009/09/099
  [arXiv:0904.4664 [hep-th]].
	 
	
\bibitem{Aldazabal:2013sca} 
  G.~Aldazabal, D.~Marques and C.~Nunez,
  Class.\ Quant.\ Grav.\  {\bf 30}, 163001 (2013)
  doi:10.1088/0264-9381/30/16/163001
  [arXiv:1305.1907 [hep-th]].
	
\bibitem{Hohm:2010pp} 
  O.~Hohm, C.~Hull and B.~Zwiebach,
  JHEP {\bf 1008}, 008 (2010)
  doi:10.1007/JHEP08(2010)008
  [arXiv:1006.4823 [hep-th]].
	
\bibitem{Hohm:2014xsa} 
  O.~Hohm and B.~Zwiebach,
  JHEP {\bf 1411}, 075 (2014)
  doi:10.1007/JHEP11(2014)075
  [arXiv:1407.3803 [hep-th]].
	
\bibitem{Marques:2015vua} 
  D.~Marques and C.~A.~Nunez,
  JHEP {\bf 1510}, 084 (2015)
  doi:10.1007/JHEP10(2015)084
  [arXiv:1507.00652 [hep-th]].
  
\bibitem{Hohm:2011dv} 
  O.~Hohm, S.~K.~Kwak and B.~Zwiebach,
  JHEP {\bf 1109}, 013 (2011)
  doi:10.1007/JHEP09(2011)013
  [arXiv:1107.0008 [hep-th]].
  
\bibitem{Sen:1991zi}
A.~Sen,
Phys. Lett. B \textbf{271}, 295-300 (1991)
doi:10.1016/0370-2693(91)90090-D


\bibitem{Polchinski:1995mt} 
  J.~Polchinski,
  Phys.\ Rev.\ Lett.\  {\bf 75}, 4724 (1995)
  doi:10.1103/PhysRevLett.75.4724
  [hep-th/9510017].
\bibitem{Meessen:1998qm} 
  P.~Meessen and T.~Ortin,
  Nucl.\ Phys.\ B {\bf 541}, 195 (1999)
  doi:10.1016/S0550-3213(98)00780-9
  [hep-th/9806120].

\bibitem{Gross:1986iv} 
  D.~J.~Gross and E.~Witten,
  Nucl.\ Phys.\ B {\bf 277}, 1 (1986).
  doi:10.1016/0550-3213(86)90429-3
	
\bibitem{Grisaru:1986vi} 
  M.~T.~Grisaru and D.~Zanon,
  Phys.\ Lett.\ B {\bf 177}, 347 (1986).
  doi:10.1016/0370-2693(86)90765-3
\bibitem{Freeman:1986zh} 
  M.~D.~Freeman, C.~N.~Pope, M.~F.~Sohnius and K.~S.~Stelle,
  Phys.\ Lett.\ B {\bf 178}, 199 (1986).
  doi:10.1016/0370-2693(86)91495-4


\bibitem{Policastro:2008hg} 
  G.~Policastro and D.~Tsimpis,
  Class.\ Quant.\ Grav.\  {\bf 26}, 125001 (2009)
  doi:10.1088/0264-9381/26/12/125001
  [arXiv:0812.3138 [hep-th]].
\bibitem{Liu:2013dna} 
  J.~T.~Liu and R.~Minasian,
  Nucl.\ Phys.\ B {\bf 874}, 413 (2013)
  doi:10.1016/j.nuclphysb.2013.06.002
  [arXiv:1304.3137 [hep-th]].
\bibitem{Minasian:2015bxa} 
  R.~Minasian, T.~G.~Pugh and R.~Savelli,
  JHEP {\bf 1510}, 050 (2015)
  doi:10.1007/JHEP10(2015)050
  [arXiv:1506.06756 [hep-th]].
  
	
\bibitem{Razaghian:2018svg} 
  H.~Razaghian and M.~R.~Garousi,
  Phys.\ Rev.\ D {\bf 97}, 106013 (2018)
  doi:10.1103/PhysRevD.97.106013
  [arXiv:1801.06834 [hep-th]].
  
 
  
\bibitem{Fukuma:1999jt} 
  M.~Fukuma, T.~Oota and H.~Tanaka,
  Prog.\ Theor.\ Phys.\  {\bf 103}, 425 (2000)
  doi:10.1143/PTP.103.425
  [hep-th/9907132].

\bibitem{Maharana:1992my} 
  J.~Maharana and J.~H.~Schwarz,
  Nucl.\ Phys.\ B {\bf 390}, 3 (1993)
  doi:10.1016/0550-3213(93)90387-5
  [hep-th/9207016].
 
\end{thebibliography}
\end{document}